# Aortic Wall Stiffness Depends on Ultrasound Probe Pressure

Marta Irene Bracco[1,2], Jonas Peter Eiberg[3,4,5], Ulver Spangsberg Lorenzen[3], Stephane Avril[1], Laurence Rouet[2]

1. Mines Saint-Étienne, University Jean Monnet, INSERM, Sainbiose, Saint-Étienne, France

2. Philips Research, Paris, France

3. Department of Vascular Surgery, Rigshospitalet, Copenhagen, Denmark

4. Department of Clinical Medicine, Faculty of Health and Medical Sciences, University of Copenhagen, Copenhagen, Denmark

5. Copenhagen Academy of Medical Education and Simulation (CAMES), Copenhagen, Denmark

Corresponding Author: Stephane Avril, avril@emse.fr

ABSTRACT

OBJECTIVE: Assessing the biomechanical behaviour of the aortic wall in vivo can potentially improve the diagnosis and prognosis of patients with abdominal aortic aneurysms (AAA). With ultrasound, AAA wall stiffness can be estimated as the diameter change in response to blood pressure. However, this measurement has shown limited reproducibility, possibly influenced by the unknown effects of the highly variable ultrasound probe pressure. This study addressed this gap by analyzing time-resolved ultrasound sequences from AAA patients.

METHOD: Two-dimensional ultrasound sequences were acquired from AAA patients, alternatively applying light and firm probe pressure. Diameter variations and stiffness were evaluated and compared. An in silico simulation was performed to support the in vivo observations.

RESULTS: It was found that half of the AAAs were particularly responsive to the probe pressure: in these patients, the cyclic diameter variation between diastole and systole changed from 1% at light probe pressure to 5% at firm probe pressure and the estimated stiffness decreases by a factor of 6.3. Conversely, the other half of the patients were less responsive: diameter variations increased marginally and stiffness decreased by a factor of 1.5 when transitioning from light to firm probe pressure. Through numerical simulations, we were able to show that the rather counter-intuitive stiffness decrease observed in the responsive cohort is actually consistent with a biomechanical model.

CONCLUSION: We conclude that probe pressure has an important effect on the estimation of AAA wall stiffness. Ultimately, our findings suggest potential implications for AAA mechanical characterization via ultrasound imaging.

INTRODUCTION

Ultrasound imaging (US) is key in surveillance programs for patients with abdominal aortic aneurysms (AAA). Specifically, abdominal US scanning enables the clinician to measure the maximum antero-posterior (AP) diameter of the aneurysm and to take clinical decisions based on the estimated risk of rupture (1,2). In general, AAA larger than 5.0/5.5 cm (female/male) are considered for treatment, and AAA smaller than this threshold will continue US surveillance. Nonetheless, it has been established that 5-10 % of ruptures occur in AAA with a diameter below this established threshold, and, additionally, 40 % of AAAs, classified as high risk based on the diameter, do not experience rupture (3–5). To augment the existing maximum diameter criterion, supplementary assessments of aortic elastic properties have been suggested, including the aortic stiffness, which is defined as the difference between systolic and diastolic blood pressures divided by the relative aortic diameter change $\Delta D/D_{DIAS}$, where $\Delta D$ is the diameter change between diastole and systole and $D_{DIAS}$ is the diastolic diameter (6–9). Given that aortic wall degeneration during AAA development alters the tissue's mechanical properties, stiffness has been proposed as a measure of such degeneration, and may indicate the severity of the disease (10).

Non-invasive US imaging can be employed for the quantification of the aortic wall stiffness (11–15). The earliest and most straightforward approach involves capturing the cyclical motion of the aortic wall by acquiring 2D time-resolved B-mode US, or cine-loops, and subsequently applying a speckle tracking algorithm to determine $\Delta D/D_{DIAS}$ (16). Combining this measurement with brachial pressure provides a global estimate for aortic stiffness. Although initially promising, this method suffers from low reproducibility and low predictive ability, hindering clinical translation (6,17). In an early study the AAA stiffness was assessed in two separate sessions two weeks apart, with two observers each time (6). Relatively high intra-observer and inter-observer variability of $\Delta D/D_{DIAS}$ and stiffness were found, suggesting that scanning conditions can influence biomechanical estimations.

A recent study reported that probe pressure (PP) is a source of bias in US morphological estimations, which might contribute to the low reproducibility of maximum diameter measurement itself (18–20). PP refers to the force exerted by the sonographer via the transducer by pushing on the patient abdomen. Considering these previous findings, it is hypothesized that the probe pressure can affect not only morphological estimations, but also stiffness estimations.

The aim of this study was to quantify the influence of PP on US-based estimations of aortic wall stiffness. For that, a specifically tailored in vivo protocol was developed to measure $\Delta D/D_{DIAS}$ under conditions of both low and high PP. Stiffness values were subsequently calculated using a standardized formulation. To further validate and provide insight into our in vivo observations, an in silico model was employed. By addressing these questions, we aim to gain a better understanding into aortic wall biomechanics in the context of US-based mechanical quantifications.

MATERIALS AND METHODS

To investigate the effects of ultrasound PP on the aortic wall stiffness, in vivo two-dimensional (2D) US cine-loops from AAA patients were collected in a clinical setting and subsequently analyzed with a semi-automatic software. Moreover, an in silico computational simulation was performed.

*2.1* In Vivo Analysis

2D US cine-loops were acquired in the outpatient clinic of Rigshospitalet Copenhagen. Patients with an infrarenal AAA following US surveillance were included prospectively and consecutively from the COpenhagen Aneurysms CoHort (COACH), thoroughly described in (21). Patients were included voluntarily and after informed consent. The study was approved by the Danish Regional Research Ethics Committee (record number H-20001116). The exclusion criteria included aorto-iliac aneurysms, suprarenal AAA, and prior aortic or iliac artery reconstruction.

2.1.1    Acquisition Protocol

The non-fasting patients were scanned with a Philips EPIQ scanner in supine position with a hand-held 2D probe (C5-1) with a spatial resolution ranging between 0.20 mm/pixel and 0.26 mm/pixel, and a temporal resolution ranging between 32 Hz and 40 Hz. First, an US compatible gel pad with an initial thickness of 1.5 cm, was placed on the patient abdomen (22). The maximum AP diameter plane was then manually located by the sonographer. Finally, two 10-second 2D US cine-loops were acquired in breath-hold conditions. The first acquisition was performed by applying minimal force, allowing an optimal visualisation of the AAA (light probe pressure, LPP). The second acquisition was conducted using the maximum transducer pressure that the patient could tolerate without major discomfort and that would still allow optimal AAA visualization (firm probe pressure, FPP). In both acquisitions, the PP and hand position were maintained throughout the scanning time. The patient was allowed to breathe between the two acquisitions. Brachial pressure was recorded just before each US scanning procedure. All acquisitions were performed by a medical doctor and PhD student with considerable experience in US AAA scanning, affiliated with the Department of vascular surgery.

2.1.2    Post Processing: Antero-Posterior Diameter Tracking

The US scans were processed offline with an interactive research prototype for US sequence analysis developed for the project and recently described in detail (12). In summary, the AAA wall was semi-automatically segmented and tracked with a speckle tracking algorithm. Then, the AP diameter was automatically detected, and the evolution of the AP diameter throughout the sequence was used to detect the diastolic and systolic peaks in the cine-loops. This analysis allowed obtaining diastolic and systolic AP diameters and the relative diameter variation $\Delta D/D_{DIAS}$ averaged over multiple cardiac cycles.

2.1.3    Post Processing: Gel Pad And AAA Depth Measurement

The measurement of the gel pad thickness and the AAA depth, i.e. the distance between the patient abdomen and the anterior AAA wall, were assessed at the centerline of the scanning plane. For display purposes, we generated pseudo M-mode images using the central scanning line of the convex B-mode sector to visualize the impact of probe pressure. For each frame, a vertical line built from the intensities of pixels lying in the center of the B-mode sector was plotted (**Figure 1**). The top surface of the gel pad (T) was detected as the first pixel with a nonzero grey value starting from the top of the image. Similarly, the bottom surface of the gel pad (B) was detected as the first non-zero value below the top surface. The distance between T and B was considered as the gel pad thickness, while the distance between B and the anterior wall (A) was considered as the AAA depth. A and P points were extracted from the 2D AP diameter tracking.

*2.1.3    Post Processing: Stiffness*

The arterial stiffness was defined as the dimensionless quantity $\beta = ln(P_{SYS}/P_{DIAS})/(\Delta D/D_{DIAS})$, where $P_{SYS}$ and $P_{DIAS}$ were obtained from the brachial pressure measurement. Such definition expresses the arterial strain as the fractional (or relative) diameter change (7).

*2.2 In silico analysis*

A finite element model (FEM) was created to simulate aortic mechanics during an US scanning procedure. A commercial software (ABAQUS Inc.) was employed for modelling the geometry, meshing and simulating the loads and boundary conditions.

2.2.1 Idealized geometry

The abdominal aorta was modelled as a cylinder with a diameter of 40 mm (similar to an AAA under surveillance but below the critical diameter of 50 mm), a length of 100 mm and a constant thickness of 2 mm. The surrounding tissues were modelled as a cylindrical body with a radius of 90 mm and a length of 100 mm. The aortic segment was put in contact with the wall of the abdominal aorta. In addition, a square (30 mm side) was extruded at a distance of 4 mm from the inner AAA wall to represent the vertebral spine. A depiction of the model cross section is shown in **Figure 2A**. The probe surface was defined as the part of the abdominal surface in contact with the US probe. Specifically, it had a rectangular shape of 70 mm x 20 mm, representing the probe footprint, and was located on the anterior part of the tissues, as shown in **Figure 2B**.

2.2.2 Meshing

The aortic segment was meshed with 1386 linear 4-node shell elements (*S4R* type, approx. size of 3 mm). The surrounding tissues were meshed with 36620 8-node hexahedral elements with hybrid formulation (*C3D8H* type, approx. size of 7.5 mm). The portion of the surrounding tissues in contact with the aorta was re-meshed with an approximate size of 3 mm.

2.2.3 Boundary Conditions

The aortic segment and the surrounding tissues were put in contact via tie constraint, i.e. no relative motion was allowed. The superior and inferior faces of the surrounding tissues were constrained in the longitudinal direction. In addition, the inner surface representing the interface with the vertebral spine was fully constrained.

2.2.4 Material Models

The aortic wall tissue was modelled as a nonlinear fiber-reinforced hyperelastic anisotropic material following the formulation by Holzapfel, Gasser and Ogden, also known as the HGO model (23,24). The values of the material constants, defining the AAA wall mechanical behavior, are summarised in **Table 1**.

These values were previously estimated for the abdominal aorta of healthy elderly individuals, matching the AAA patients group (25). A plane stress conditions was considered for the shell model of the aorta.

The surrounding tissues were modelled as a homogeneous incompressible Neo-Hookian hyperelastic material (26) with a shear modulus of 20 kPa, as in (27).

2.2.5 Loads

Blood pressure values were determined within a physiological range (120 mmHg / 80 mmHg) and adjusted for abdominal aortic pressure using correction coefficients of 1.05 (systole) and 0.78 (diastole) (28). The simulation procedure is schematized in **Figure 3**. In the first simulation step, a diastolic blood pressure was applied to the inner surface of the aortic model. In the second step, the US PP applied onto the patient abdomen was simulated by a uniform pressure of either 2 kPa (LPP) or 30 kPa (FPP), taken from the pressure values measured in the cine-loops. Finally, in the third step, the blood pressure was increased to simulate the systolic phase.

RESULTS

3.1 In vivo analysis

All the cine-loops from the 10 patients (9 males, 1 female), aged from 63 to 86, were successfully processed. The main results for all the patients are presented in **Figure 4**. The application of FPP resulted in a decrease of the systolic AP diameter between 1.02 mm and 6.84 mm, depending on the patient. Conversely, an increase of the relative diameter change was observed in all cases except one, where a small decrease of $\Delta D/D_{DIAS}$ was detected. By looking at the graph, based on these results, the cases can be split in two classes: five patients showing a net increase in $\Delta D/D_{DIAS}$ (from a mean of 1.1 % to 5 %), referred to as the 'responsive cohort' and the other five showing small increase or decrease (from a mean of 1.1 % to 1.4 %), referred to as the 'passive cohort'. The estimated stiffness β decreased on average by a factor of 6.3 in the responsive cohort, and by a factor of 1.5 in the passive cohort.

**Table 2** summarizes the results found in these two cohorts. In addition, pseudo M-mode images were produced to visualize the effect of the PP on the sequence dynamics. Two representative cases are reported in **Figure 5**. The cyan and pink overlay plots indicate the gel pad thickness over time. The gel pad deformation in LPP and FPP conditions can be appreciated. In **Figure 5A**, belonging to the responsive cohort, a cyclic motion in the bottom part of the gel pad can be observed in FPP condition. The blue and red overlays show the anterior and posterior aortic walls. The responsive case shows increased cyclic motion in the aortic wall, while in the passive case (**Figure 5B**), such phenomenon is not visible. In addition, the effect of increased US PP on surrounding tissues thickness, i.e. the aortic depth, can be appreciated. For the sake of conciseness, only two representative cases are presented here. The rest of the M-mode sequences can be found in the supplementary material.

3.2 In silico analysis

Although the FEM simulations were performed in three dimensional space, the results are presented here only for the cross sectional view of the aorta to allow direct comparison with the image data. **Figure 6** shows the cross-sectional view of the aortic segment, oriented as **Figure 2A**. The contours of the diastolic configuration obtained after the application of US PP (end of Step 2), are also plotted. The systolic configuration reached after Step 3 is plotted as a color map showing the displacement field.

Finally, in order to compare the simulation and acquisition outputs, the AP diameter evolution in two representative in vivo sequences (**Figure 7A** and **B**, corresponding to **Figure 5A** and **B**) and in the in silico simulations (**Figure 7C**) are plotted. To aid comparison, the simulations results were plotted in multiple cardiac cycles (three per PP condition), mimicking the physiological time course of aortic expansion. The dashed lines in the plots allow to appreciate the AP diameter reduction due to LPP and FPP.

DISCUSSION

The primary objective of this study was to investigate how the pressure from an US transducer affects aortic wall mechanics. Specifically, the in vivo mechanical response of AAAs to varying levels of PP was characterized. Counterintuitively, it was observed that applying FPP can lead to an increased distensibility of the aortic wall in the AP direction. Consequently, the present findings suggest that ultrasound PP can artificially decrease the measured stiffness of the aorta.

Previous works have shown that PP exhibits considerable variability among different experienced sonographers (18,22). The magnitude of PP is susceptible to many factors, including the composition of abdominal tissues and the presence of bowel gas (6). The present study reveals a notable reduction of up to 7.9 mm in the anterior-posterior (AP) diameter when transitioning from LPP to FPP conditions. These findings align with those presented in a previous study Ghulam et al, wherein the authors established that PP represents a potential source of error in measurements of the AP diameter (18). In addition to confirm these previous findings, the present study is the first to date showing the effects of US PP on dynamic diameter variation assessments and stiffness estimations.

A prior work focused on assessing the reproducibility of stiffness quantifications and ΔD measurements (6). It reported an intra-observer coefficient of variation of ΔD measurement up to 27 % when scans were separated by two weeks, along with an inter-observer coefficient of variation of 20 %. These values indicate a reasonable range of measurement variations but led to substantial fluctuations in indirect assessments, such as stiffness. In the present study, higher variations in both ΔD and stiffness were found (up a mean of +273 % and -84 % respectively, in the responsive cohort), most likely due to our specific objective to emphasize changes in PP rather than conducting a reproducibility analysis. In fact, in the present study, the sonographer received explicit instructions to achieve the two extreme pressure conditions during image acquisitions.

Within the present findings, two distinct response patterns to PP were delineated and categorized: in half of the cases, referred to as the "responsive cohort," a discernible augmentation in relative diameter change was observed when transitioning from LPP to FPP conditions. In contrast, the remaining half, referred to as the "passive cohort" exhibited a minor increase in relative diameter change. This dynamic response is reflected in stiffness calculations, particularly within the responsive cohort, where the AAA wall appeared up to 6.3 times stiffer under LPP compared to FPP. Conversely, in the passive cohort, stiffness was overestimated by up to 1.5 times under LPP conditions.

Instinctively, one may erroneously predict that, due to the mechanical constraint caused by an increased opposing external force, a larger stiffness would be measured when a firm probe pressure is applied, whereas, here, the opposite has been observed. To elucidate this behavior, the aortic mechanics during an US exam was investigated in silico. FEM was employed to simulate the LPP and FPP conditions acting on an idealized aortic segment by introducing surrounding tissues. The quantification of PP values was derived from measurements of the gel pad and was converted into pressure values using the method outlined by Svendsen et al. (22). Accordingly, LPP was set at 2 kPa and FPP at 30 kPa. The results of the FEM simulations corroborated the responsive cohort's behavior, namely, an increase in relative diameter variation (decrease of stiffness) following the application of FPP. Two primary factors can explain this behavior: the flattening of the geometry due to increased external pressure and the nonlinear nature of the elastic response within the aortic tissue. The flattening effect can be interpreted with the Laplace law. When an external firm pressure is applied, the cross-sectional shape of the artery becomes elliptical, as illustrated in **Figure 6**. Consequently, the radius of curvature increases in the anterior and posterior walls while decreasing in the lateral walls. According to Laplace's law, in a vessel of constant thickness, with constant blood pressure, a higher radius of curvature corresponds to an increased wall stress. This increased stress is a function of the material's elasticity and strain, which can be expressed through relative diameter variation $\Delta D/D_{DIAS}$. Therefore, $\Delta D/D_{DIAS}$ is expected to increase as the radius of curvature increases, e.g. in the case of FPP. Nonetheless, in this study we have considered an idealized geometry. However, it is known that patient-specific geometry can substantially influence the mechanical behavior of an AAA (29). Therefore, future investigations should take into account the patient-specific anatomical features for a more comprehensive analysis.

On the other hand, both groups of patient have similar reductions of the AP diameter due to PP, as shown in Figure 4. Therefore, the geometrical changes due to PP cannot alone explain the effects of PP on the aortic stiffness. Note that we have also measured a higher gel pad deformation in the passive cohort (37 %) than in the responsive cohort (18 %), indicating that the surrounding tissues might be stiffer in this population, hence limiting the aortic distensibility. Thus, we submit that the mechanical properties of the surrounding tissues play a major role in the individual response to PP. Moreover, the effects of PP on aortic stiffness are likely to be related to the nonlinearity inherent to the aortic wall tissue, characterized by a linear relationship between tissue stiffness and stress. This relationship has been known for almost 60 years (30) and researchers have discovered that the linear coefficient between stiffness and stress may vary significantly between subjects (31). As a matter of fact, the alteration of mechanical forces acting on the aorta might induce a change in the elastic response of the wall. To investigate this last hypothesis further, more simulations are needed to compare the AAA wall tissue mechanical response to external PP when either linear or nonlinear material properties are prescribed. These findings may potentially open new opportunities for in vivo characterization of the aortic nonlinear elastic behavior.

The presented work unveils a previously unknown phenomenon that sparks some light on the effect of US PP on AAA biomechanics. However, such finding needs to be fully understood in order to become exploitable in a clinical setting. Notably, even though the analyzed sample was small, two different groups of patients were clearly observable. In a future study, our preliminary findings should be confirmed on a larger population. In addition, it could be interesting to perform the same acquisition with more than two values of probe pressure per patient, in order to explore the relationship between ΔD and PP.

Finally, we have noticed that, in the responsive cohort, FPP can induce a motion in the bottom part of the gel pad, indicating that the aortic expansion affects the surrounding tissues. In future work, the impact of surrounding tissues material properties on both the gel pad deformation and the AAA mechanics should be investigated.

CONCLUSIONS

Through the combination of in vivo and in silico techniques, the presented study has elucidated that PP has a significant influence on both diameter variations and in vivo estimated stiffness of the aortic wall. The in vivo analysis on 10 patients lead to the conclusion that some AAAs show a dramatic stiffness decrease when a firm PP is applied. We have proposed an interpretation based on geometrical changes and material nonlinearity, underpinned by the results of a FEM analysis. Our hypothesis needs further validation through comprehensive clinical investigations encompassing larger patient cohorts and patient-specific computational simulations. A deeper understanding of this somewhat unanticipated phenomenon will hopefully open new possibilities for aortic mechanical characterization and AAA rupture risk estimation.

*Acknlowledgements* This project has received funding from the European Union's Horizon 2020 research and innovation program under the Marie Skłodowska-Curie grant agreement No 859836.

*Conflict of Interest Statement* MIB, JPE and LR are affiliated with Koninklijke Philips N.V.

*Data Availability Statement* The authors confirm that the data supporting the findings of this study are available within the article and its supplementary materials.

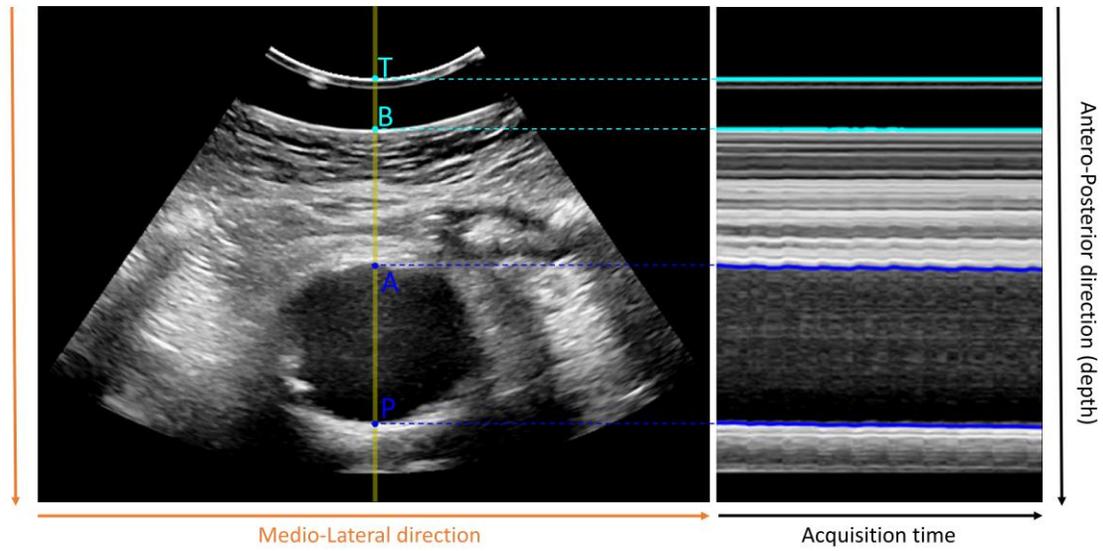

**Figure 1**. The procedure to obtain a pseudo M-mode US image is shown. The centerline of the B-mode image (left) is extracted at every frame of the sequence and stacked horizontally into the pseudo M-mode image (right). The motion of the top (T) and bottom (B) points of the gel pad (cyan), and the anterior (A) and posterior (P) walls of the AAA (blue) is displayed as overlay.

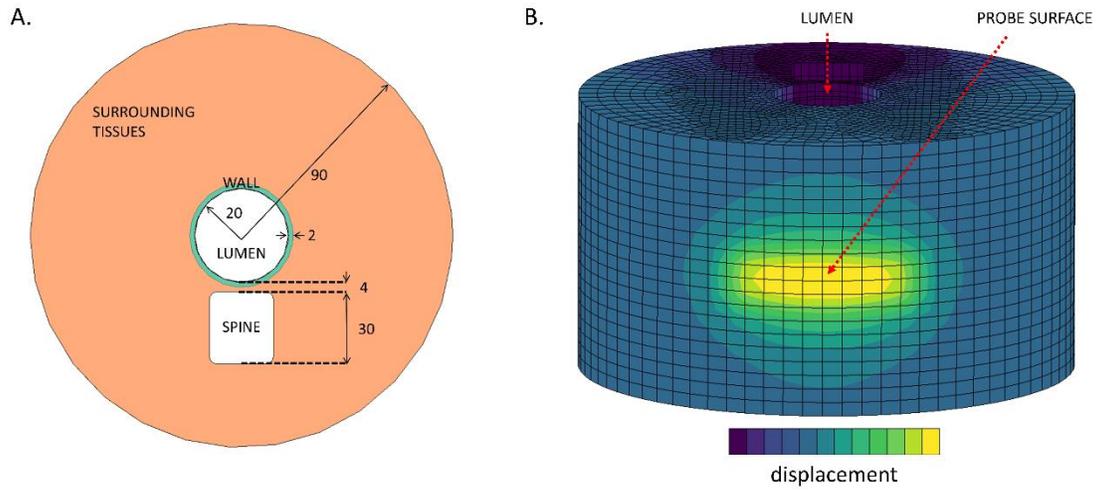

**Figure 2**. Idealized model of the abdominal aorta embedded in surrounding tissues. The cross section of the model (A) comprises of the lumen, the spine, the aortic wall (green) and the surrounding tissues (salmon). The dimensions are reported in mm. The solid model (B), obtained by extruding (A) is shown together with the mesh and the displacement map (starting from 0) obtained by applying an arbitrary probe pressure on the probe surface.

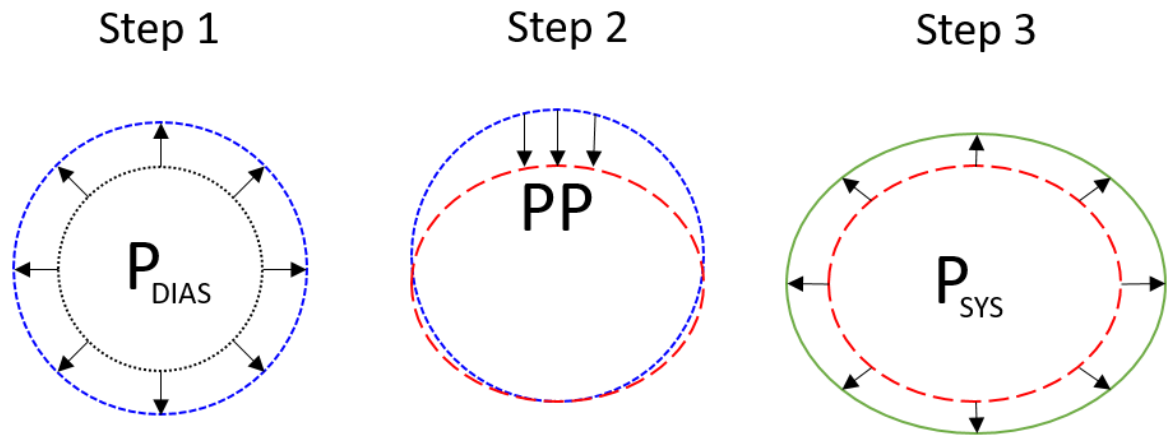

**Figure 3**. Schematic explanation of the simulation procedure. In Step 1, the diastolic blood pressure onto the abdomen. Finally, in Step 3, the blood pressure is increased to the systolic value (PSYS).

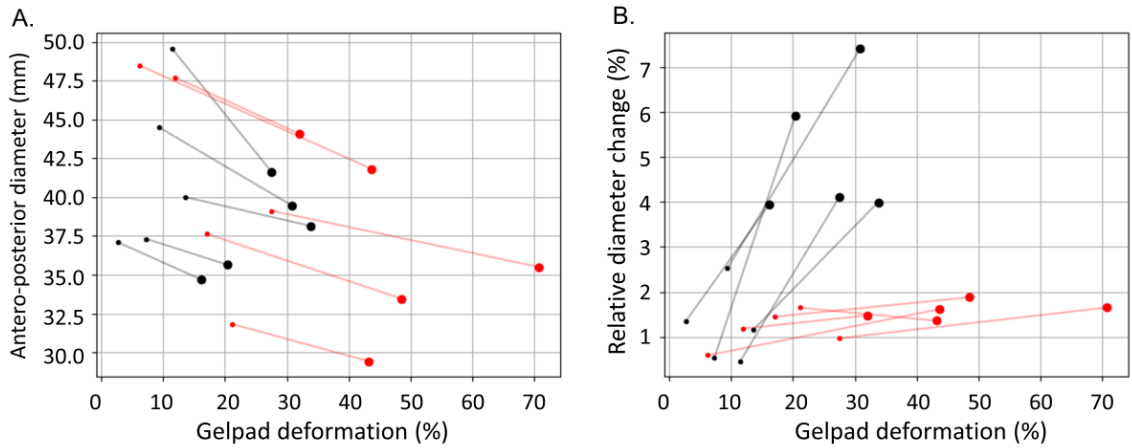

**Figure 4**. Results from the ten patients. The systolic antero-posterior (AP) diameter (A) and the relative diameter change (B), computed as the diameter variation with respect to the diastolic AP diameter, are plotted against the relative deformation of the gel pad. The latter is computed with respect to the original gel pad thickness of 15 mm. The small dots represent the light probe pressure condition, while the big circles indicate firm probe pressure. The sequences plotted in black belong to the cohort of patients that was judged 'responsive' because their diameter change is highly affected by the application of probe pressure. The other cases are highlighted in red and they belong to the less responsive 'passive' cohort.

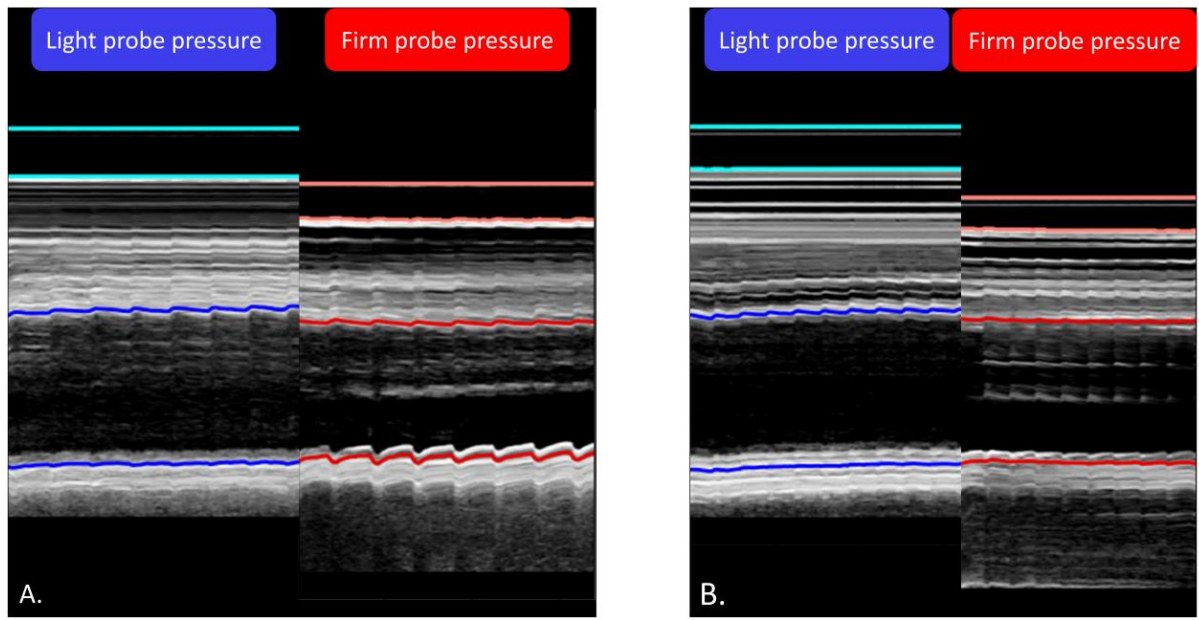

**Figure 5**. 2 cases (A) and (B) showing different reaction to increased probe pressure. For each case, two pseudo M-mode images compare the light and the firm probe pressure conditions (left side/right side). In order to reflect the fixed position of the spine, the posterior wall in the sequence obtained with light probe pressure (blue) is aligned to the corresponding posterior wall from the firm probe pressure sequence (red). Case (A): the motion of the posterior wall is increases with probe pressure. Case (B) no change in the posterior wall motion between the LPP and FPP conditions. Because the posterior wall of the aorta is attached to the spine, it cannot move during the acquisition. Thus, although a motion in the posterior wall can be measured in the US sequence, that motion is actually happening in the anterior wall. This fact is due to the inherent nature of handheld US technique, where the imaging probe itself is displaced by the pulsating tissues.

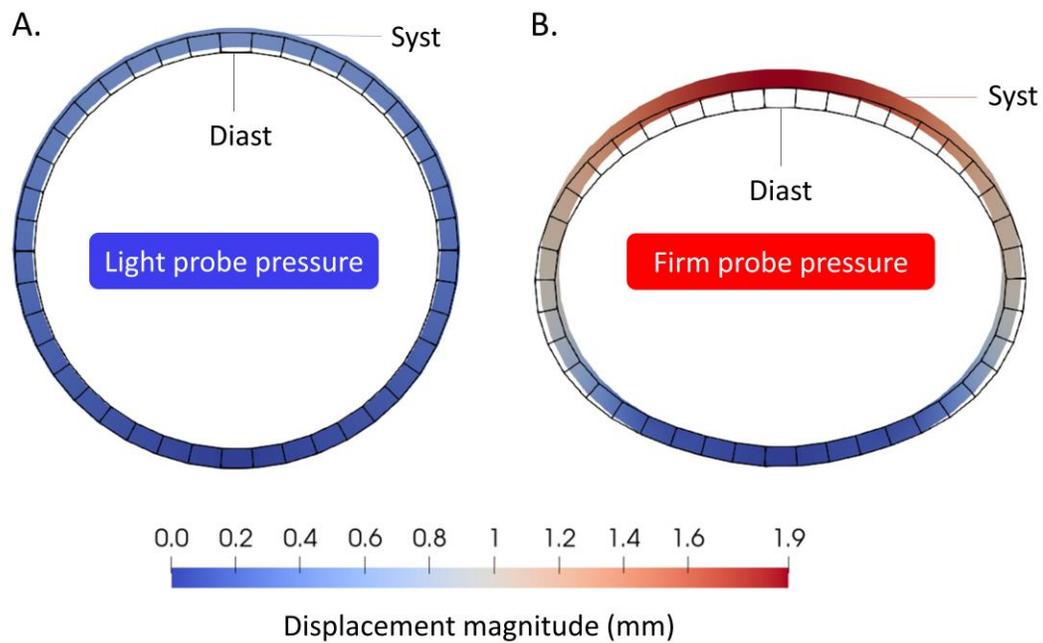

**Figure 6**. The results of the two finite element simulations are presented for the mid cross-section of the aortic wall model. A is for the light probe pressure case (left) and B for the firm probe pressure (right), the diastolic and systolic profiles of the vessel are presented after the application of the probe pressure (Step 3 in Figure 3). The wireframe shows the diastolic configuration. The color maps showing the displacement magnitude (in mm) from the diastolic to the systolic configuration are plotted on the systolic configuration.

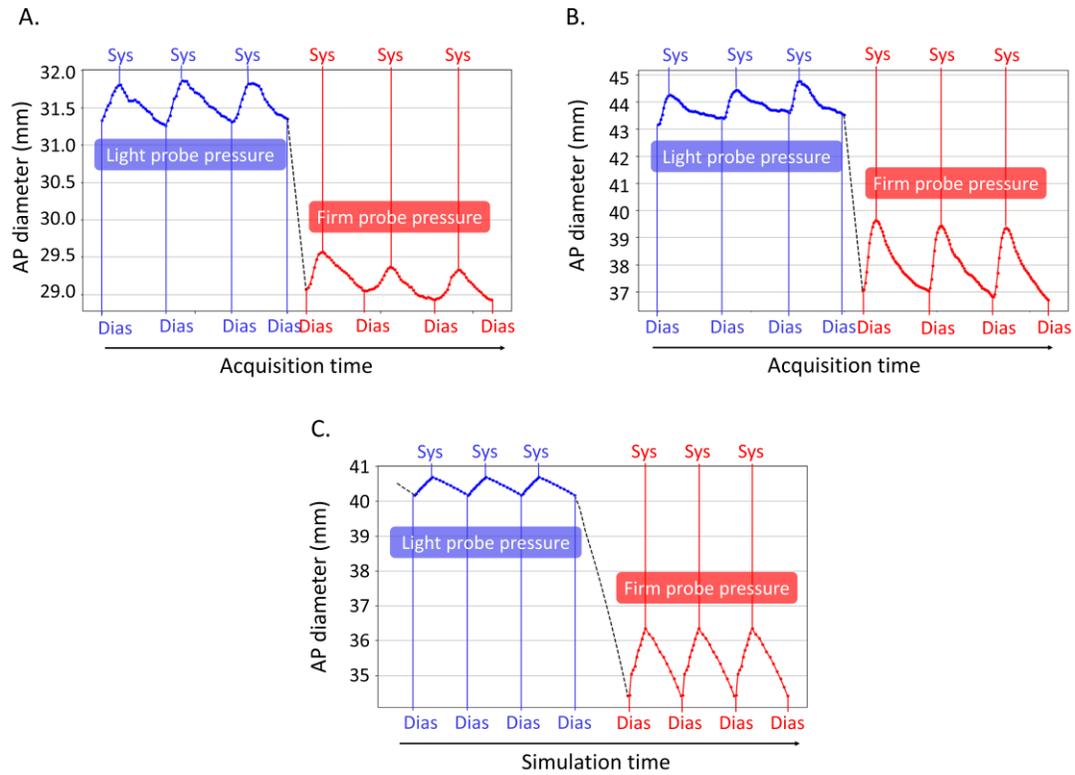

**Figure 7**. The Antero-posterior diameter evolution is shown for the in vivo sequences: one reactive case (A) and one passive case (B) same cases as in Figure 5. To compare in vivo and in silico methods, the simulation results (C) were plotted for three cycles. In each plot, the diameter evolution at light probe pressure (blue) is followed by the one at firm probe pressure (red). The black dashed line represents the diameter reduction due to the compressive force exerted by the probe. Please note that the y axes are not in the same scale.

**Table 1**. Material Constants for the aortic wall material model following the formulation by (23,24). Values taken from (25). Mechanical parameters: $C_{10}$ = initial stiffness (kPa), $k_1$ (MPa) and $k_2$ *(adimensional)* = stiffness parameters, K (GPa) = bulk modulus. Structural paramenters: κ (adimensional) = collagen fibers dispersion coefficient, Θ (°) = collagen fibers orientation angle.

| $C_{10}$ (kPa) | $k_1$ (MPa) | $k_2$ | κ | Θ (°) | K (GPa) |
|---|---|---|---|---|---|
| 100.9 | 4.07 | 165.55 | 0.16 | 48.4 | 75.5 |

**Table 2.** Quantifications from the light probe pressure (LPP) and firm probe pressure (FPP). The top and bottom parts of the table refer to the cases plotted in Figure 4 in black and red respectively. The quantities are reported with the notation: mean (standard deviation). From left to right, we report the systolic and diastolic diameters ($D_{sys}$, $D_{dias}$, in mm), the absolute diameter variation ($\Delta D$, in mm), the stiffness (adimensional), the gel pad thickness ($GP_{thick}$, in mm), and the surrounding tissues thickness ($ST_{thick}$, in mm).

|  | $D_{sys}$ (mm) | $D_{dias}$ (mm) | $\Delta D$ (mm) | Stiffness $\beta$ | $GP_{thick}$ (mm) | $ST_{thick}$ (mm) |
|---|---|---|---|---|---|---|
| Responsive cohort | | | | | | |
| LPP | 41.70 (4.75) | 41.19 (4.7) | 0.49 (0.32) | 63.42 (45.66) | 13.67 (0.57) | 30.47 (5.27) |
| FPP | 37.92 (2.52) | 36.14 (2.40) | 1.83 (0.51) | 9.98 (2.75) | 11.14 (0.98) | 22.71 (4.0) |
| change | - 9 % | - 12 % | 273 % | - 84% | - 18 % | - 25 % |
| Passive cohort | | | | | | |
| LPP | 40.96 (6.32) | 40.45 (6.34) | 0.45 (0.11) | 36.71 (16.3) | 12.49 (1.1) | 44.03 (7.35) |
| FPP | 36.86 (5.4) | 36.28 (5.32) | 0.58 (0.1) | 24.64 (7.28) | 7.84 (1.92) | 33.99 (8.3) |
| change | - 10 % | - 10 % | 28 % | - 33 % | - 37% | - 22 % |